%% file: main.tex
\documentclass[manuscript]{acmart}
\usepackage{kotex}
\usepackage{amsmath}
\usepackage{svg}
\usepackage{enumitem}

\input{header1.tex}
\begin{document}
\input{header2.tex}

\section{Introduction}

Recent advancements in artificial intelligence, graphics, and robotics are accelerating the realization of artificial embodied agents like social robots, virtual avatars, and digital humans. The embodied artificial agent is one of the promising interaction mediums beyond current voice-only assistants, and their non-verbal behavior is a key difference from the existing mediums. When we interact with agents having a human form factor, we naturally expect them to do social behaviors as humans do, which is often called ``social affordance''~\cite{shu2017learning}. This non-verbal behavior has been proven to be effective in many human-human interaction and human-agent interaction studies~\cite{mcneill1992hand, burgoon1990nonverbal, sauppe2014robot, bremner2011effects, wilson2017hand}. Proper gesticulation of agents helps to reveal agents' intention~\cite{sauppe2014robot}, to make listeners concentrate~\cite{bremner2011effects}, and to build a rapport with humans~\cite{wilson2017hand}.

It is not trivial to make plausible non-verbal behavior for the agents. Typical ways to realize social behaviors are keyframe animation by artists or capturing human actors' motion~\cite{menache2000understanding}. Both methods give high-quality motions but there is a scalability issue due to their high production cost. Experienced animation artists are needed for the keyframe animation, and the motion capture method requires expensive capturing devices and skilled actors. Thus, these gesture authoring methods are mostly used in movies, structured games (not open-world games), small-sized interaction applications where virtual characters' or agents' utterances are limited. 

There is a need for a cheaper solution. The agents that interact with humans freely, like social robots, virtual assistants, and characters in interactive games, have diverse utterances, so it is not feasible to create gestures for all the utterances using the expensive aforementioned methods. The cheapest way to generate social behavior is automatic generation; it does not require a human effort at the production time. For facial expressions, there are audio-driven facial motion generation studies~\cite{karras2017audio, thies2020neural} and a commercial toolkit is also available for the same purpose~\cite{CrazyTalk8}. A body gesture is another important nonverbal behavior, and an automatic generation of body gestures is more challenging than a generation of facial motion due to the lack of large-scale gesture datasets. There are two automatic gesture generation toolkits---Virtual Humans Toolkit~\cite{hartholt_all_2013} from USC and NAOqi~\cite{naoqi} from SoftBank Robotics; however, both toolkits are not widely used because they are dependant on a specific platform and heavily relies on pre-described word--gesture association rules. Also, the gesture quality of the automatic gesture generation toolkits is not high enough, showing repetitive and monotonic motion and unnatural transitions between gestures. Recent data-driven gesture generation studies improved gesture qualities, but are still far behind human-level gestures~\cite{geneaworkshop}. 
Another limitation of the data-driven gesture generation is that it cannot imply a designer's intention. Even if generated gestures are as natural as human gestures, they might not be what a designer wants. 


\begin{figure}[t]
  \centering
  \includegraphics[width=0.7\linewidth]{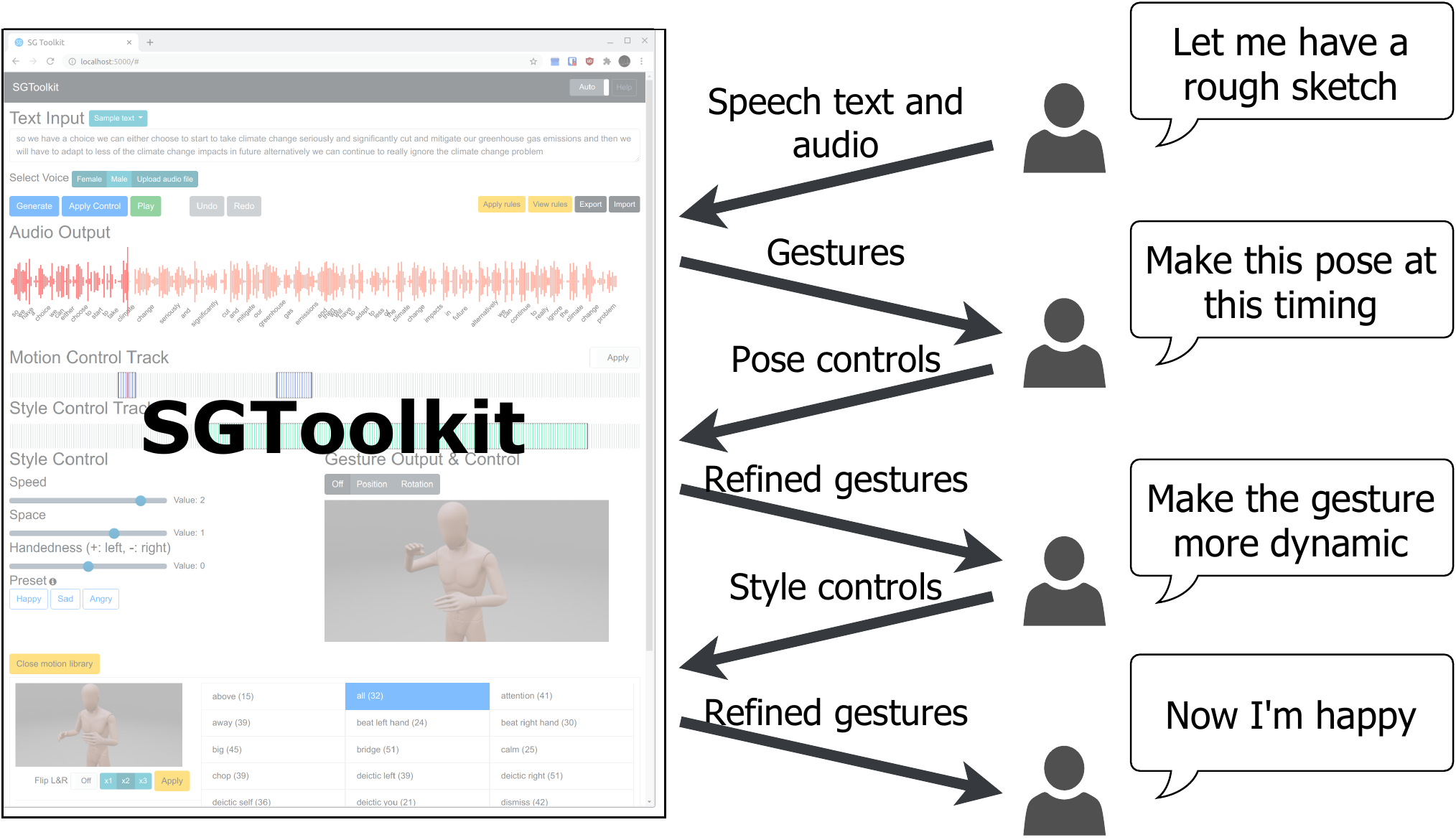}
  \caption{An exemplary usage flow of the SGToolkit. A user first reviews the automatically generated gestures (a rough sketch), and then adds pose and style controls until they are satisfied with the refined gestures.}
  \label{fig:flow}
\end{figure}

In this paper, we propose SGToolkit, a novel gesture generation toolkit that enables designers or developers to create quality gestures with less effort. The toolkit is designed to have higher gesture quality than automatic gesture generation methods and lower production cost than a manual keyframe animation method. 
We first interviewed three experts in potential fields, where the toolkit can be used, and elicited the requirements for the gesture generation toolkit (Section \ref{sec:requirements}). The key idea of the new gesture generation toolkit is combining automatic gesture generation and manual controls. Figure \ref{fig:flow} shows the workflow of the proposed toolkit. The toolkit first generates gestures from speech automatically, and then users add control elements to refine the gestures. The toolkit provides fine-level pose control and coarse-level style control elements. We describe how the controls are integrated into a neural generative model in Section \ref{sec:model}, and how the neural network was trained on a human gesture dataset in Section \ref{sec:training}. Note that we focus on upper-body gestures which include head, arm, and torso movements, and that both pose control input and synthesized output were represented as 3D joint positions.
Section \ref{sec:SGToolkit} describes how the toolkit was developed after a preliminary user study with its first prototype.
And we evaluated the usability of the toolkit and the quality of the generated gestures (Section \ref{sec:user study}). 


\section{Related Works} \label{sec:related works}

\subsection{Gesture Generation Toolkits}

There are two representative gesture authoring toolkits publicly available. One is the Virtual Humans Toolkit from USC~\cite{hartholt_all_2013} which is for conversational virtual characters; it provides a complete set of toolkits including a dialogue engine, a character animation engine~\cite{thiebaux2008smartbody}, and a non-verbal behavior generator~\cite{lee2006nonverbal}. The non-verbal behavior generator is based on a rule-based gesture generation method. Rules mapping from words to predefined unit gestures were specified in a configuration file, and users can modify and expand the configuration. The other toolkit is the NAOqi framework~\cite{naoqi} for NAO and Pepper robots. NAOqi also uses a rule-based gesture generator similar to the Virtual Humans Toolkit. Several unit gestures are embedded in the robots, and the robots make gestures according to words currently speaking. 

The existing toolkits have the common disadvantages that they are platform dependant (the SmartBody virtual characters~\cite{thiebaux2008smartbody} for the Virtual Humans Toolkit and the robots from the SoftBank Robotics for the NAOqi framework) and generated gestures are often repetitive and not matching to speech because both toolkits generate gestures only from a limited modality of speech text. When input speech does not have words specified in the gesture generation rule, the existing toolkits generate static or random motion, which is not relevant to speech, and, without considering speech audio, it is not possible to synthesize gestures aligning to speech acoustically, for instance, beat gestures. In addition, the previous toolkits were for automatic gesture generation, so there are no interactive authoring or controlling functions. To our best knowledge, the proposed SGToolkit is the first interactive gesture generation toolkit supporting fine- and coarse-level controls.

Other relevant studies are to build a markup language for gesture authoring~\cite{cassell2004beat, kopp2006towards}. Although these studies introduced useful tools for manual gesture creation, a bunch of unit gestures is still needed and a user should be aware of the unit gestures to write down markup codes. 

\subsection{Automatic Gesture Generation Methods} \label{sec:gesture methods}

There are lots of studies for automatic gesture generation for embodied conversational agents. The studies have improved gesture generation quality by adapting advanced machine learning methods and considering multimodality. Automatic gesture generation methods can be categorized into a rule-based approach~\cite{kopp2006towards, cassell2004beat, marsella2013virtual} and a data-driven approach~\cite{yoon2019robots, kucherenko2020gesticulator, huang2014learning, levine2010gesture}. The existing toolkits we introduced above use the rule-based approach because its implementation is straightforward and it is easy to extend and reuse generation rules. 
Unlike the rule-based approach requiring human-generated rules, the data-driven approach learns how to make gestures from a large corpus of human gestures; the studies in this category are widely spanned according to motion representation (discrete unit gestures vs. continuous motion) and speech context considered (speech audio, speech content, interlocutor, etc.). Early studies focused on building gesture generation rules automatically from human demonstrations~\cite{kipp2005gesture, huang2014learning}, and later, the researchers proposed methods generating raw motion data not using predefined unit gestures~\cite{yoon2019robots, kucherenko2019analyzing, ferstl2019multi, ginosar2019gestures}. Also, the research has evolved to use multimodal speech context to reflect speech information as much as possible~\cite{kucherenko2020gesticulator, yoon2020speech}.


The model in the present paper is based on the data-driven method, and we added controllability on top of that. Both speech audio and text are considered in the model. 

\subsection{Controllable Gesture Generation}

Manipulating gesture motion can be done in two different ways. The first one is using post-processing for existing gestures. EMOTE~\cite{chi2000emote} introduced a computational method to manipulate speed, range, and path of character motion based on Laban Movement Analysis~\cite{laban1971mastery}. Hartmann et al.~\cite{hartmann2005implementing} proposed a similar post-processing method for speech gestures of virtual characters. While the two studies focused on manipulating overall motion statistics, Liao et al.~\cite{liao2020speech2video} inserted desired poses to an existing motion to generate a new motion; the inserted poses were blended into the existing motion. These post-processing methods directly manipulate motions, so there is the advantage that the updated motion is predictable. The disadvantage is that sophisticated motion blending functions are necessary and users should tweak blending parameters manually to get realistic motion.

Another way to control gesture motion is inputting control intents to a generation model, and the model considers the intents during motion generation processes. Alexanderson et al.~\cite{alexanderson2020style} proposed a style-controllable gesture generation model. Style control vectors were inputted to the neural network model, and it was demonstrated that the model generates gestures following the inputted style intents. Also, speaker identities were used to generate stylized gestures reflecting inter-person variability~\cite{yoon2020speech, Ahuja2020Style}. A style embedding space was trained from a large corpus of gestures of different speakers in~\cite{yoon2020speech}. A style disentanglement was studied in~\cite{Ahuja2020Style}.

Our model adapted the later approach and extended it to control both details of gestures and rough styles, which was not suggested before. 



\section{Requirements Finding Interview} \label{sec:requirements}

To find out the requirements for a gesture authoring tool that enables an efficient authoring process and can create human-like gestures, we conducted interviews with experts in three different domains related to gesture creation: video games, massive open online courses (MOOC), and social robots. Interviewee 1 is a character motion researcher from one of the world's largest video game studios. Interviewee 2 is the CTO of a MOOC company. The company uses virtual characters with gesture motions in its educational video materials. Interviewee 3 is a robotics researcher; his team developed a new android robot and demonstrated the robot at exhibitions and performances.

We interviewed each interviewee for an hour. At the beginning of the interview, we presented four existing gesture authoring methods: keyframe animation, motion capturing, rule-based automatic gesture generation (NAOqi framework~\cite{naoqi}), and deep learning-based automatic gesture generation~\cite{yoon2019robots}. We explained their concepts and showed videos for actual gesture authoring processes and results. Next, we asked several open-ended questions. The questions were 1) general opinions about the four methods, 2) requirements for a gesture authoring tool, and 3) appropriate autonomy level of gesture creation (fully manual to fully automatic). 

Since the experts' domains were all different, we could get insights into a broad field. We list the requirements that we extracted from the insights in the following. First, all the interviewees agreed that automatic gesture generation and manual editing are both necessary. Interviewee 2 particularly favored automatic gesture generation because substantial effort was needed to create character motion for a large collection of educational materials. However, the interviewees commented that manual motion editing is still needed. Interviewee 1 said editing the generated motions is required because the quality of automatically generated gestures is not as good as what a designer can create manually. He noted that automatically generated gestures could be used as rough sketches for designers. Interviewee 2 commented that a virtual character should be able to make different gestures according to background contents (e.g., deictic gestures indicating a paragraph on a blackboard the character is currently explaining). To accomplish this, the toolkit should provide a way to set specific poses at specific time frames.

Second, the tool should control not only the motions but also more abstract things, such as emotion and styles. Interviewee 3 said gestures do not always contain a concrete meaning; sometimes they express abstract feelings. He said it would be great if the tool can create gestures based on emotions. Interviewee 2 noted that different gesture styles are required based on the audience. For example, content for children needs a lot of gestures. In contrast, content for company executives does not have many gestures. 

Third, the tool should be versatile. We found that there is a need for creating co-speech gestures in various fields. Accordingly, the gesture generation tool should be usable on various platforms, and the tool should consider non-designers, such as educators (Interviewee 2). 

To meet both requirements for automatic generation and manual editing, we decided to create the gesture authoring toolkit that creates gestures automatically from speech and provides ways to modify the generated gestures. The tool should satisfy the following requirements specifically. 
\begin{itemize}
    \item (\verb|Req.1|) The toolkit should allow a user to alter the gesture motion by controlling one or more poses at specific time frames. 
    \item (\verb|Req.2|) The toolkit should support modifying abstract aspects of gestures, such as style.
    \item (\verb|Req.3|) The toolkit should output gestures in a general format and provide platform-independent APIs so that users in different fields can use them.
\end{itemize}

\section{Controllable Gesture Generation Model} \label{sec:model}

We introduce a controllable gesture generation model that accommodates manual controls from a user on top of automatic gesture generation results. The model is based on the previous state-of-the-art gesture generation model~\cite{yoon2020speech}, and we integrated two types of manual controls---fine-level human pose controls and coarse-level style controls---to increase the quality of output gestures and controllability. In the following subsections, we describe the model architecture and two control elements in detail.

\subsection{Overall Architecture}

We selected the speech gesture generation model proposed by Yoon et al.~\cite{yoon2020speech} as a base model, which aimed to generate gestures automatically. The base model was designed to generate a sequence of human poses from speech text, speech audio, and speaker identity; here, speaker identity was used to indicate different styles of gestures. The base model consisted of three encoders for each input modality to understand speech context and a decoder to synthesize a sequence of upper-body human poses. Each human pose $\hat{p}$ at time $i$ was generated by the model $G_{base}$ as follows: $\hat{p}_i = G_{base}(a_i, w_i, spk)$ where $a_i$, $w_i$, and $spk$ are the encoded feature vectors for speech audio, speech words, and speaker identity, respectively. The encoders and the decoder were deep neural networks (CNN for audio and RNN for words and poses), and the whole network was trained in an end-to-end fashion. 

\begin{figure}[t]
  \centering
  \includegraphics[trim=0cm 11.5cm 14.5cm 0cm,clip,width=0.6\linewidth]{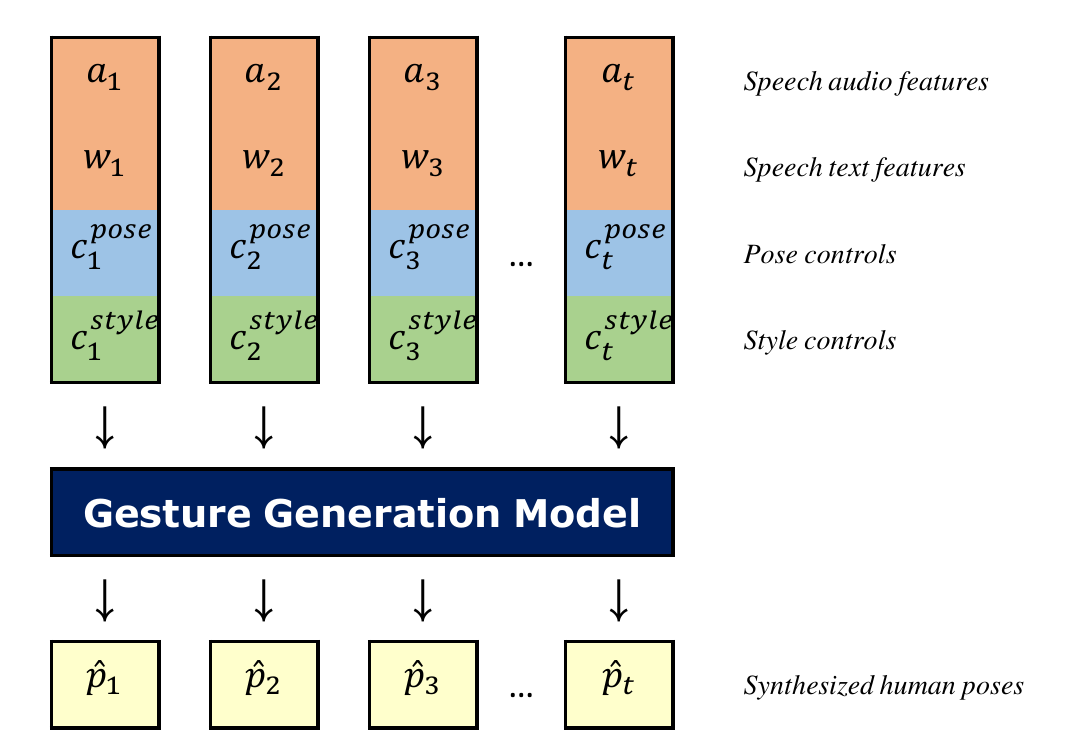}
  \caption{Model architecture. The model generates a sequence of human poses from speech context and user-inputted controls. All the speech feature vectors, controls, and output motion are in the same time resolution.}
  \label{fig:model}
\end{figure}

We revised the base model to accommodate pose and style controls. Both controls were represented as a collection of control vectors (details will follow in the next subsections). The control vectors were concatenated as a conditioning information to the speech feature vectors: $\hat{p}_i = G(a_i, w_i, c^{pose}_i, c^{style}_i)$ as shown in Figure \ref{fig:model}. $c^{pose}_i$ and $c^{style}_i$ are pose and style controls and they are in the same time resolution to the speech features and output gestures. As the style control is introduced, we removed the speaker identity in the base model because its purpose, generating stylized gestures, is the same as the style control. 


We fixed $t$, the number of generated poses, in Figure \ref{fig:model} to 30 which is 2 s long (motion was in 15 fps throughout the paper). This does not mean that the proposed toolkit is only able to generate gestures up to 2 s. The model concatenates several generation results to make a long motion sequence as the base model did in the original paper. Please see Appendix \ref{appendix:long} for the details.

\subsection{Fine-level Human Pose Control}

The pose controls are for specifying desired poses at certain time frames (\verb|Req.1|). For example, you might want a deictic gesture pointing to the lower-right corner as saying ``... look at the first shelf on the right side ...'' or bowing motion for greeting words. These gestures are less likely to be generated by data-driven gesture generation models because of the scarcity of deictic and iconic gestures and cultural biases in a training dataset. In the proposed model, a user inputs desired poses or motion to the model, then the model generates plausible gestures while following the inputted pose controls as much as possible.

The pose controls are defined as a sequence of desired poses concatenated with mask bits: 
\begin{equation} \label{eq:pose}
  c^{pose}_{i=1...t} = (\tilde{p}_i, mask^{pose}_i),
\end{equation}
\begin{equation}
  mask^{pose}_i=
    \begin{cases}
      1, & \text{if}\ \tilde{p}_i \text{ is specified} \\
      0, & \text{otherwise}
    \end{cases}
\end{equation}

\noindent where $\tilde{p}$ is a user-specified control pose in the same data representation as the output pose $\hat{p}$. The binary mask bits indicate regions where control poses were specified. 
For example, if a user wants the virtual character to make a deictic gesture pointing right at the 10th time frame, the user might set the 10th frame of $\tilde{p}$ to a pose pointing right. The remaining frames should be uncontrolled. In this example, $mask^{pose}_{i=10}=1$ and $mask^{pose}_{i \neq 10}=0$.
There is no restriction on the number of controls and their continuity; users may input a single control pose or gesture motion in a varying length. If there is no specified pose control, the model runs like a fully automatic gesture generator. 

\subsection{Coarse-level Style Control}

The pose control is an effective element to generate particular gestures, but it could be too restrictive and requires an effort to input control poses. Also, as the requirement mentions (\verb|Req.2|), the tool has to be able to change abstract factors of gestures. Therefore, we added the style control which is less restrictive and easier to input. 
The style control is for manipulating overall motion statistics, namely \textit{speed}, \textit{space}, and \textit{handedness}. \textit{Speed} is for the temporal property of gestures and \textit{space} for the spatial property. We also used \textit{handedness} as a means to show agents' personalities. The motion statistics \textit{speed} and \textit{space} were used as style parameters to represent emotional states in \cite{chi2000emote, hartmann2005implementing}, and \textit{handedness} were considered in~ \cite{alexanderson2020style}.

We defined the style control as follows:
\begin{equation} \label{eq:style}
  c^{style}_{i=1...t} = ({speed}_i, {space}_i, {handedness}_i, mask^{speed}_i, mask^{space}_i, mask^{handedness}_i).
\end{equation}

\noindent $Speed$ represents the average speed of all the joints, $space$ means the distance between the left and right hands, and $handedness$ represents the relative speed of the left and right hands. Similar to the pose control, we used binary mask bits that indicate the use of style elements. More specifically, the style vectors were calculated at the training stage as follows:
\begin{align}
    \label{eq:style1}
    {speed}_i &= \frac{1}{w} \sum_{k\in\mathcal{J}} \sum_{j=i-w/2}^{i+w/2} \lvert p_j^k - p_{j-1}^k \rvert , \\
    \label{eq:style2}
    {space}_i &= \frac{1}{w} \sum_{j=i-w/2}^{i+w/2} \lvert p_j^{left wrist}-p_j^{right wrist} \rvert , \\
    \label{eq:style3}
    {handedness}_i &= 
    \begin{cases}
      speed^L / speed^R - 1, & \text{if}\ speed^R > speed^L \\
      1 - speed^R / speed^L, & \text{otherwise,}
    \end{cases}
\end{align}

\noindent where $p_j^{joint name}$ means the position of $joint name$ for $j$th time frame, $speed^L=\frac{1}{w} \sum_{j=i-w/2}^{i+w/2} \lvert {p}_j^{left wrist}-{p}_{j-1}^{left wrist} \rvert$, and $speed^R$ is defined similarly with $p^{right wrist}$. $\mathcal{J}$ is the set of the upper-body joints used in the paper. The motion statistics were calculated on a region centered at $i$ with the fixed window size of $w=30$ with a stride of 1. All the style values were normalized with the mean and standard deviation on the training dataset and clamped to be in range [-3, 3]. 

This style control mechanism, concatenating desired style control signals to speech feature vectors, was originally introduced by Alexander et al.~\cite{alexanderson2020style}. In the original paper, one style element was used per model, so it was not possible to manipulate multiple styles at the same time. In the present paper, we used three styles of speed, space, and handedness, and integrated them into a single model.

\section{Training and Validation} \label{sec:training}

We trained the controllable gesture generation model on the TED dataset~\cite{yoon2019robots, yoon2020speech} which contains 97 h of TED talk videos and 3D human poses extracted automatically. We split the dataset into training (80\%), validation (10\%), and test sets (10\%). The test set was only used for the user study and qualitative result figures. We followed the same training scheme in~\cite{yoon2020speech}. The speech context encoders and gesture generator in the model were trained from scratch except using the pretrained word embeddings~\cite{bojanowski2016enriching} for speech text understanding.

The key difference to the base model is the use of the pose and style controls. We simulated the controls, which are supposed to be specified by human users at the test time, from the reference human motions in the training dataset. For pose controls, a part of human motion $p_{i=1...t}$ in a training sample was used. This subsequence is sampled to be in a varying length between 1 and $t$ and its position is selected randomly to simulate diverse pose controls. The style controls were also simulated from the style statistics of the reference motion data. During the training, we randomly dropped all the controls to expose the model to the scenario of automatic gesture generation without pose and style controls. One of three style elements was also dropped randomly to simulate diverse style combinations.


\subsection{Objective Function}

The model was trained by the following objective function:
\begin{equation} \label{eq:loss}
  L_{total} = \alpha \cdot L_{Huber}(p, \hat{p}) + \beta \cdot L_{GAN}(\hat{p}) + \gamma \cdot L_{style}(s, \hat{s})
\end{equation}

\noindent where $s=style(p)$ and $style(\cdot)$ is the style calculation function (Eq. \ref{eq:style1}--\ref{eq:style3}), which is differentiable. $L_{Huber}$, a mixed form of L1 and L2 losses, minimizes differences between the generated motion $\hat{p}$ and reference human motion $p$ for the same speech. $L_{GAN}$ is the adversarial loss~\cite{goodfellow2014generative} to assess how much the generated gestures look like real human motions. We added $L_{style}$ that is defined as L1 loss between style vectors $s$ from $p$ and $\hat{s}$ from $\hat{p}$. Although there is no explicit loss term for pose controls, the model was trained to follow pose controls because copying inputted pose controls to output poses is the easiest way to minimize $L_{Huber}$. In other words, $L_{Huber}$ drives the model to work like an identity function that passes pose controls into output poses for controlled regions. $L_{style}$ also helps the model to follow style controls in the same way as the pose control.

The weights $\alpha$, $\beta$, and $\gamma$ were determined experimentally to 500, 5, and 0.05. The model was trained during 80 epochs and it took about 8 h on an RTX 2080 Ti.

\subsection{Qualitative Validation}

\begin{figure}[t]
  \centering
  \includegraphics[trim=0cm 4.5cm 10cm 0cm,clip,width=0.9\linewidth]{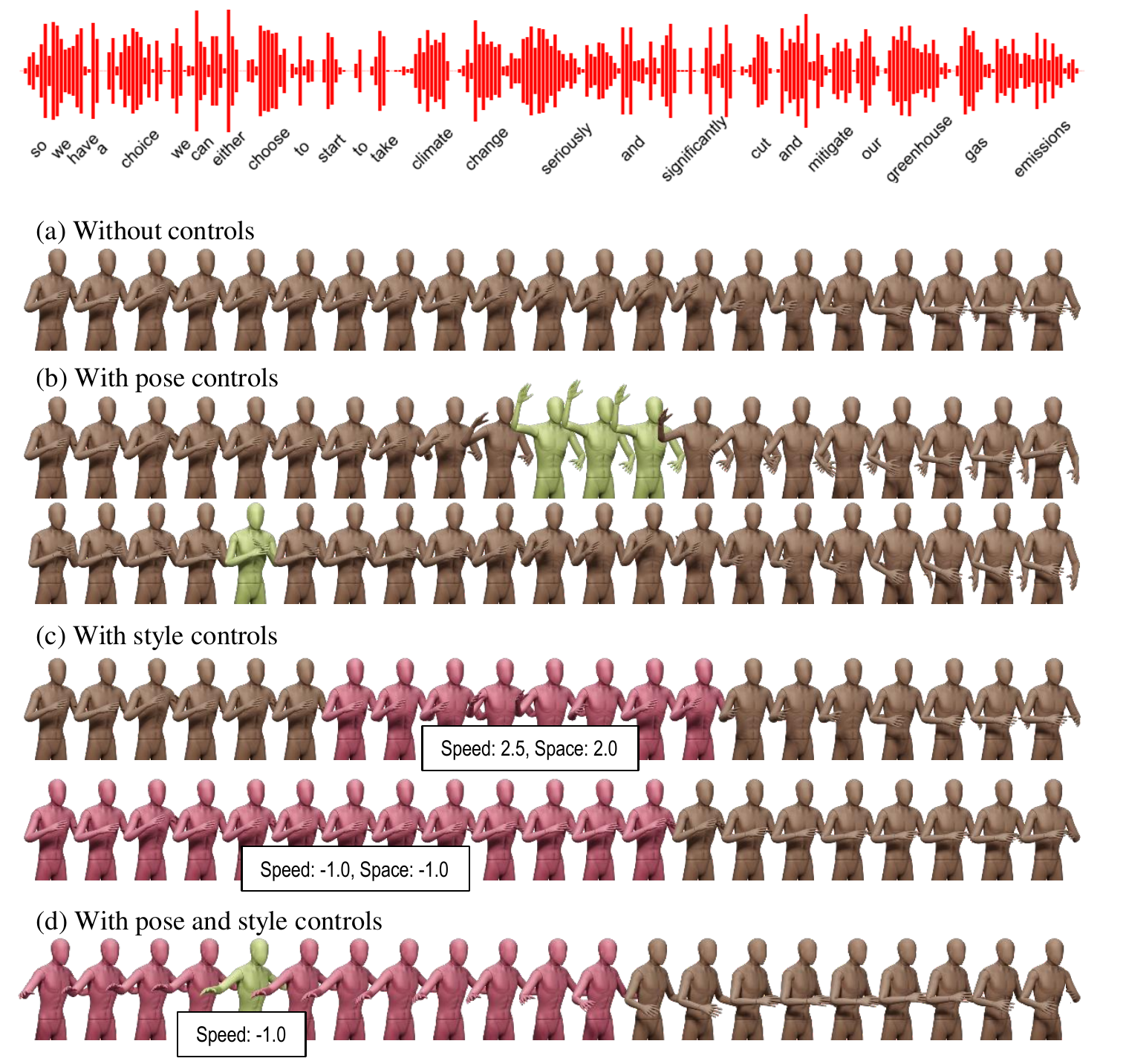}
  \caption{Gesture generation results (a) without controls, (b) with pose controls (two examples), (c) with style controls (two examples), and (d) with pose and style controls for the same speech input. The regions where the controls are applied are highlighted (green for pose controls and red for style controls). Speech audio, words, and motion are on the same time axis. Motion frames are sampled equidistantly for visualization. More results on the supplementary video.}
  \label{fig:results}
\end{figure}

Figure \ref{fig:results} shows the gesture generation results with and without controls for the same speech. The model was able to generate gestures without any controls, which means the model could behave like an automatic gesture generator (Figure \ref{fig:results}(a)). When the pose or style controls were given, the model generated gestures well reflecting the controls. Figure \ref{fig:results}(b) shows the results when a motion or a pose was used as pose controls. We observed smooth motion transitions before and after the pose control regions. Figure \ref{fig:results}(c) shows the results for the different style parameters. Also, the model worked as expected when both pose and style controls were applied as shown in Figure \ref{fig:results}(d). We added a single-frame pose control where two arms were stretched horizontally and set low \textit{speed} around the pose control. This resulted in gesture motion maintaining the control pose for some duration not raising or lowering arms owing to the style control of \textit{speed=-1}. Please see the supplementary video for more results.

It was interesting that the generated gestures with pose controls (Figure \ref{fig:results}(b)) are different from the gestures without controls (Figure \ref{fig:results}(a)) not only in the controlled regions but also in the non-controlled regions. For instance, the ending poses in Figure \ref{fig:results}(a) and (b) are all different even though there are no controls for the ending frames. This happens because the generation model tries to generate the most suitable motions in both controlled and non-controlled regions. In the first example in Figure \ref{fig:results}(b), when the motion of raising an arm was used as pose controls, the model continues to generate dynamic motion which is suitable to the previous gestures. 
Some users might dislike unexpected changes. In that case, users can simply add pose controls with automatically generated poses to prevent changes.


\subsection{Numerical Evaluation}

The trained model was evaluated using three numerical metrics (subjective evaluation will be followed in Section \ref{sec:user study}). First, we evaluated overall gesture generation results using Fr\'{e}chet Gesture Distance (FGD) which was introduced in~\cite{yoon2020speech} to evaluate automatic gesture generation models. FGD compares the distributions of human gestures and synthesized gestures on a latent feature space, for evaluating generation quality and diversity altogether. We also evaluated how well the model complies to pose controls $\tilde{p}$ and style controls $\tilde{s}$, which are named pose compliance score (PCS) and style compliance score (SCS) in the present paper. PCS is defined as the absolute difference between the pose controls $\tilde{p}$ and the synthesized poses $\hat{p}$; SCS is defined similarly. 
\begin{align}
    \text{PCS} &= average(\lvert \tilde{p}_i - \hat{p}_i \rvert) \text{  for } i \in [10, 15)\\
    \text{SCS} &= average(\lvert \tilde{s}_i - \hat{s}_i \rvert) \text{  for } i \in [1, t]
\end{align}
\noindent Here, we assumed 5-frame pose controls in the middle of a sequence in PCS and entire style controls in SCS. Pose controls can be inserted in any place in varying lengths, but we fixed the position and length of the pose control for the reproducible and consistent evaluation. Since a gesticulation style does not change rapidly, we used style controls for the entire frames (30 frames) in SCS.

We compared three models including the proposed one, a baseline model without pose and style controls, and a model always generating the static mean pose. The baseline model is the same as the proposed model, except that control elements of $c^{pose}$ and $c^{style}$ are excluded in the model and its training.
Table \ref{tab:numeric} shows the evaluation results.
We evaluated FGD for the scenarios where there are no controls (which means automatic gesture generation only from speech text and audio) and where pose or style controls exist. First, we found better performance (lower FGD) when pose or style controls were used in the proposed model. FGD is reduced to 0.194 with pose controls and to 0.228 with style controls; both FGD is lower than the FGD of the baseline (0.368). These are expected results since adding controls would drive the model to generate gestures similar to the reference motions. We would say that the proposed model generates better gestures than the automatic generation method when a proper pose or style control is used. When we consider only the scenario of no controls, the baseline showed a slightly lower FGD (0.368) than the proposed model (0.447). As the proposed model is trained to meet multiple objectives of automatic generation and supporting pose and style controls, the proposed model is degraded in FGD than the baseline which have fewer conditioning elements; however, the increase in FGD is 0.079 which is small compared to the FGD of the static mean pose (26.768). Note that the FGD values reported in this paper are not comparable with the values in~\cite{yoon2020speech} since the data representation is different.

The proposed model showed the PCS of 0.014 and the SCS of 0.249. The PCS of 0.014 is equivalent to the average error of 0.96 degrees in joint angles; we believe angle differences less than 1 degree are small enough. The SCS of 0.249 is also small considering the style value range of [-3, 3]. To help understand how small these values are, we reported PCS and SCS for the static mean pose (considered as a baseline performance) in Table \ref{tab:numeric} although the mean pose does not care about the controls.

\begin{table}
  \caption{Numerical evaluation results for the proposed model, the baseline model without pose and style controls, and the static mean pose. FGDs were reported for the scenarios with no controls, with pose controls, and with style controls. Lower is better for all the metrics. We report the PCS and SCS for the static mean pose to help you understand their ranges (indicated with *), although the mean pose does not consider the pose or style controls.}
  \label{tab:numeric}
  \begin{tabular}{llllll}
    \toprule
    Model & \multicolumn{3}{l}{FGD↓} & PCS↓ & SCS↓\\ 
    \cmidrule{2-4}
     & No controls & w/ pose controls & w/ style controls &  &  \\
    \midrule
    Proposed & \phantom{0}0.447 & 0.194 & 0.228 & 0.014 & 0.249 \\
    Baseline & \phantom{0}0.368 & -- & -- & -- & -- \\
    Static mean pose & 26.768 & -- & -- & 0.143* & 1.236* \\
  \bottomrule
\end{tabular}\end{table}

\section{SGToolkit} \label{sec:SGToolkit}

We developed SGToolkit, an interactive speech gesture authoring toolkit, by combining the controllable gesture generation model with user interfaces for inputting controls, animating output gestures, etc. The development was done through iterative prototyping. The first prototype was a stand-alone PC application. We conducted a preliminary user study with the first prototype, and then re-designed the toolkit and implemented a web application with a few improvements according to the suggestions raised during the preliminary user study. 

\subsection{The First Prototype}

\begin{figure}[b]
  \centering
  \includegraphics[width=0.8\linewidth]{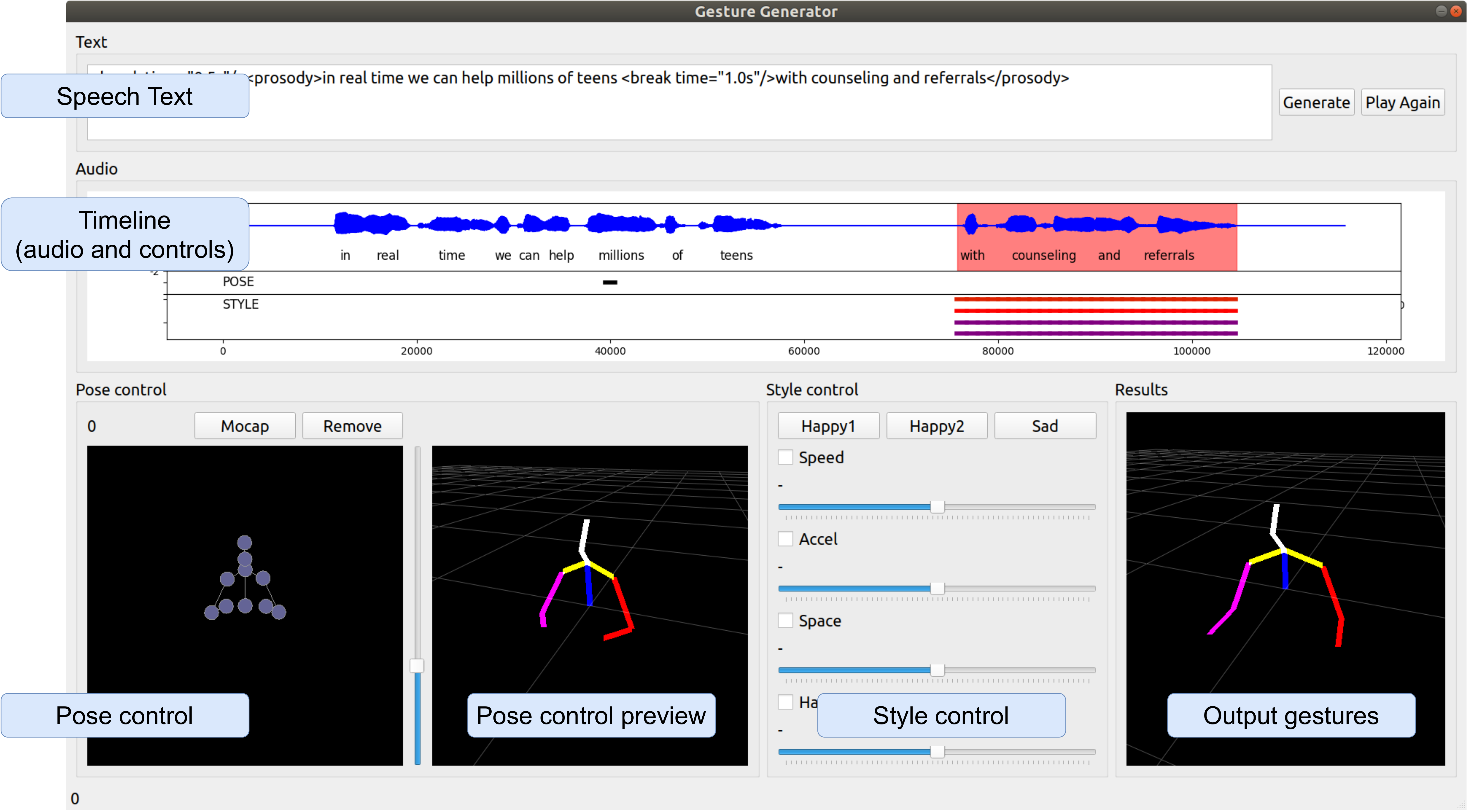}
  \caption{The first SGToolkit prototype.}
  \label{fig:1stprototype}
\end{figure}

The first SGToolkit prototype was implemented using the PyQT framework as a stand-alone PC application (Figure \ref{fig:1stprototype}). The user interface consisted of a speech text input box, timeline for audio and controls, pose control panel, style control panel, and animation panel showing output gestures. The usage flow was that: 1) input speech text and run gesture generation to get a rough sketch of the gestures, 2) add pose and style controls, and 3) run gesture generation with the controls and repeat steps 2 and 3 as many as a user wants. The timeline displays speech audio signal, speech words, pose controls, and style controls, so that users could add pose or style controls at the desired time position. Users manually set the pose controls by moving the joint positions. For the style controls, a user changes style values using sliders for four styles of speed, acceleration, space, and handedness (acceleration was excluded in the final prototype). In addition to the style control sliders, we added style presets; there were two presets of happy (speed: 2.5, acceleration: 2.0) and sad (speed: -2.5), which were determined based on the studies on expressive gestures~\cite{calvo2015expressing, knight2014expressive}. When a user adds controls, they could select a single frame by mouse clicking or a region of multiple frames by mouse dragging.

The gesture generation model required speech audio and text to generate gestures. When only speech text is available, we synthesized speech audio using Google TTS~\cite{googletts}. A forced aligner~\cite{ochshorn2016gentle} was also used to get timestamps of each word. 

It is cumbersome to craft desired poses manually, especially gestures---a sequence of poses. Thus, we added a camera-based motion capturing (mocap) function to make it easy to input pose controls. By using the mocap function, users could record their gesture motion and crop a region to input as pose controls. We integrated the existing human pose estimation method~\cite{sun2019deep, pavllo20193d}, which produces 3D human poses from RGB camera images, to the toolkit prototype.

\subsection{Preliminary User Study}

We conducted a user study with the first prototype. We recruited three graduate students from local universities and asked them to try the toolkit to generate speech gestures. Two are experienced in character animations for 3+ years and the other one has 3+ years of research experience in user interfaces. We prepared 10 sample sentences assuming a scenario of a shopping mall guide robot. One example is ``Welcome to the mall. Our mall is the largest shopping mall with over 250 shops, 8 screen cineplexes, and plenty of gourmet cafes and restaurants.'' The study took about an hour for each participant and 14 USD was paid for each.

We collected think-aloud comments while using the toolkit and conducted post-experiment interviews. All the participants commented that they liked the usage flow of getting a rough sketch by automatic gesture generation and adding controls to modify the results, and they were mostly satisfied with the generated gestures. However, there were several suggestions regarding user interfaces. 

For the pose control interface, the participants wanted a more intuitive way of manipulating gestures. This prototype let users control XY positions of joints with a 2D skeleton view and Z positions with a separate slider control, but the participants wanted to move joints more freely in a 3D space. Also, the participants said it was a little bit tedious to set the 3D position of each joint to set pose controls. The participants suggested a collection of pre-defined gestures that are commonly used to make adding pose controls easy and fast. One participant commented that separated pose views for pose controls and output gestures could be integrated. For the style control, the participants commented that the checkboxes for activating style elements are unnecessary. Style controls could be activated when a user manipulates the control sliders. In addition, there was a request for the undo and redo functions during manipulating controls, and the camera-based mocap was not completely satisfactory in terms of accuracy. 

We agreed to design more intuitive control interfaces for our final prototype. Also, we decided to drop the mocap function due to its limited accuracy.

\subsection{The Final Prototype}

\begin{figure}[t]
  \centering
  \includegraphics[width=0.45\linewidth]{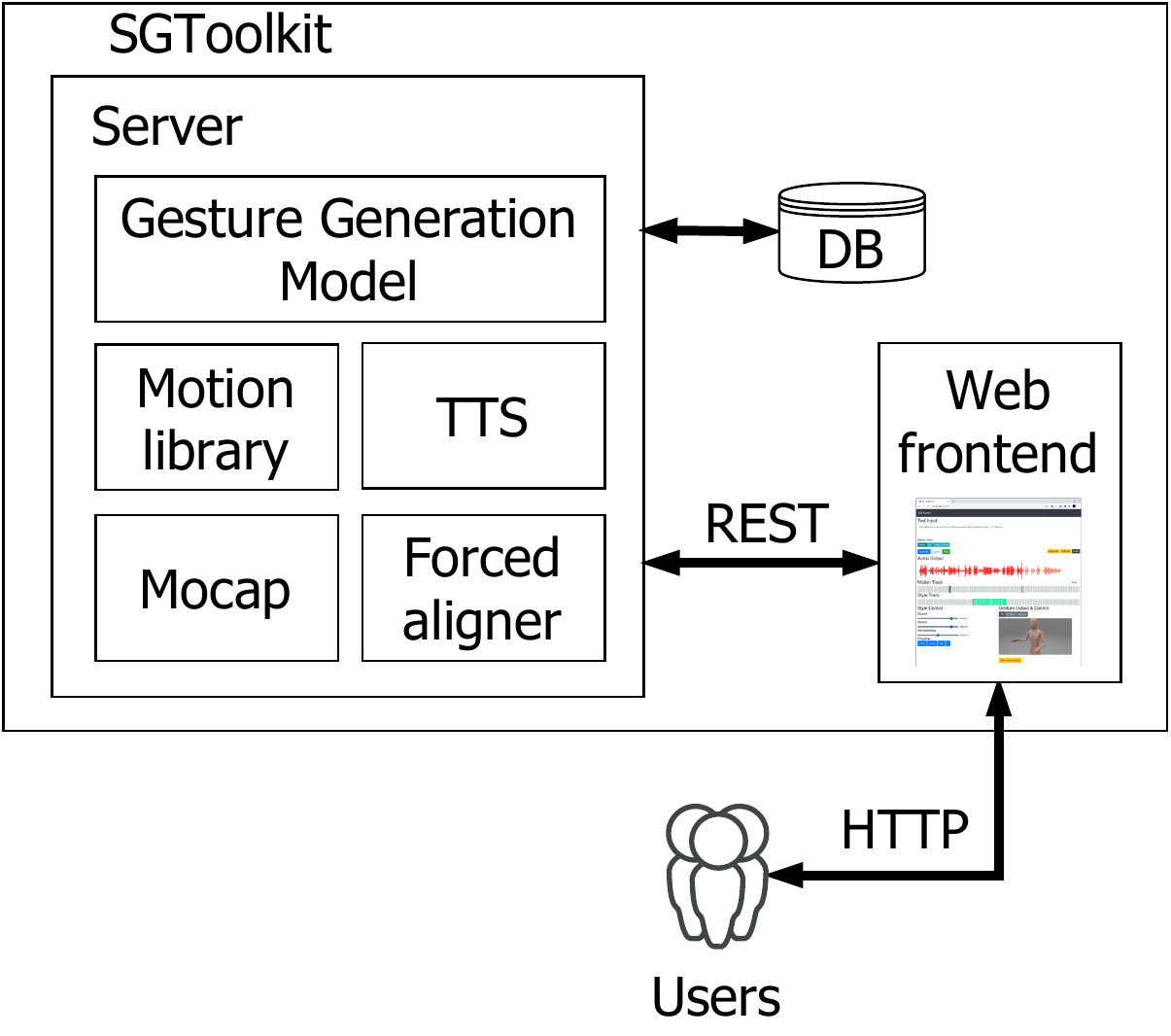}
  \qquad
  \includegraphics[width=0.45\linewidth]{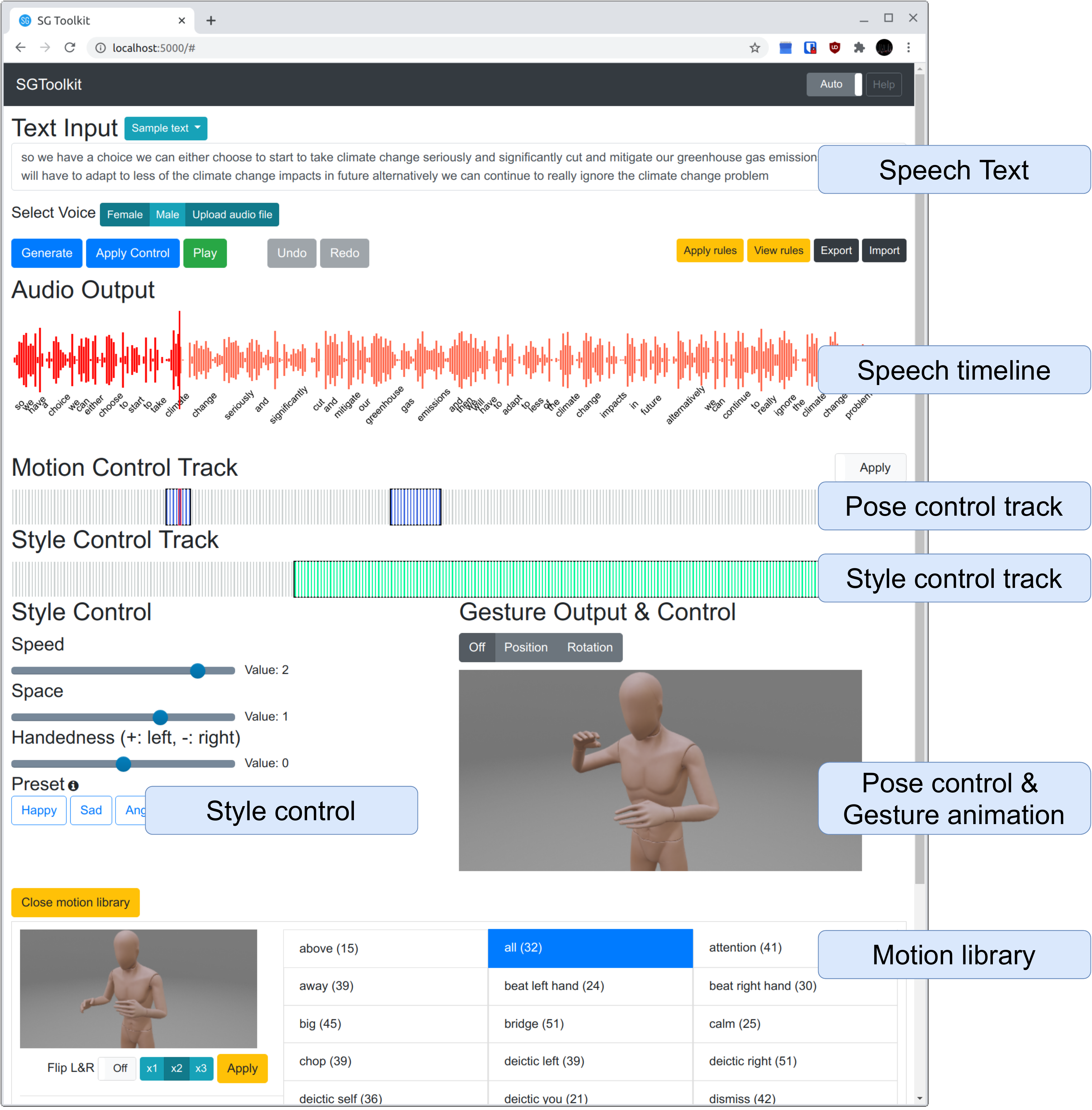}
  \caption{(left) SGToolkit components and (right) web user interface.}
  \label{fig:toolkit}
\end{figure}

We re-designed the SGToolkit according to the comments above and implemented it in a server--client structure for flexibility. We implemented the client as a web application for interoperability (running on modern web browsers supporting HTML5). Figure \ref{fig:toolkit} shows the internal components and web user interface. Users interact with web interfaces over HTTP and the web front-end communicates with the python server via REST APIs to send speech and control information and to get the generated results. We implemented only the web interface, but we followed conventional REST API designs so that other developers may implement other clients, for instance, gesture generation plugins for Unity\footnote{\url{https://unity.com}} or Robot Operating System (ROS)\footnote{\url{https://www.ros.org}} (\verb|Req.3|).

We introduced a motion library in this prototype. The motion library is a collection of unit gestures such as raising hands, pointing right, and framing gestures. Users can select one of them as pose controls instead of manual pose editing. We prepared 33 unit gestures listed in~\cite{kipp2005gesture}, and this motion library can be expanded as users add more motions. There are options for unit gestures to adjust motion speed in three levels (1x, 2x, and 3x) and to flip left and right. The unit gestures in the motion library were created using mocap and manual pose editing.


The panels of the pose controls and output gestures in the first prototype were merged into a single character animation panel. For pose controls, users manually set the 3D coordinates of each joint via joint rotations or translations with fixed bone lengths. We used a 3D mannequin character from Mixamo\footnote{\url{https://www.mixamo.com}} and Babylon.js~\cite{moreau2016babylon} for the gesture animation.

In the style control, we excluded acceleration which is largely overlapped with speed. And a history function was implemented which stores the states of controls to support undo and redo. We also made inputted controls movable along the timeline by mouse dragging. 

\section{User Study} \label{sec:user study}

We conducted a user study to compare the SGToolkit to automatic gesture generation and keyframe animation. The user study consisted of two sessions: 1) a gesture design session where participants create gestures and 2) a gesture quality evaluation session to evaluate gestures created in the previous session and gestures created automatically. In the gesture design session, the participants created gestures using the SGToolkit (named \textit{SGT} condition) and using the manual keyframe animation (\textit{MANUAL} condition). 


As the SGToolkit aims to create quality gestures with less effort, we measured both quality and efficiency in this user study. The quality and efficiency are in inverse proportion, and it is hard to measure both metrics objectively altogether. Also, in a pilot test with a few subjects, we had difficulties measuring authoring time, which is related to efficiency, because the participants tended not to end an authoring task when there is no time limit. To this end, we decided to measure efficiency and maximum achievable quality by having two scenarios where the maximal authoring time was fixed to 5 min (rushed) and 20 min (relaxed). We evaluated the two factors in separate scenarios. In the first rushed scenario, we assessed how well a participant can design gestures in a short and limited time. In contrast, for the relaxed scenario, participants had plenty of time to create gestures. The purpose of a long time was to let participants author gestures with the highest quality. The time limits were decided after a pilot study. 
A participant was assigned either one of the scenarios and authored gestures in the aforementioned conditions.


\subsection{Gesture Design}

Participants created gestures under the \textit{SGT} and \textit{MANUAL} conditions. 
For the \textit{MANUAL} condition, we implemented a keyframe animation method. When a user puts key poses to certain frames, the method fills motion in between the key poses using cubic spline interpolation for each joint, as typical keyframe animation does. The same user interface was used for both conditions to reduce the effects of user interfaces. Participants specified key poses in the same way as the pose controls---moving or rotating joints--in \textit{SGT}. The motion library, a collection of unit gestures, was also usable in both conditions. In \textit{MANUAL}, the mean pose was added as key poses at the first and last frames to prevent undesired extrapolation artifacts. Note that participants created gestures from scratch without having a rough sketch and the style controls were unavailable in the \textit{MANUAL} condition.

\subsubsection{Procedure}

The overall procedures for the two scenarios were similar. First, an experimenter explained what SGToolkit is and how to use it for 20 min. Then a participant created gestures for several sentences under two conditions. Lastly, the participant completed a survey, followed by a short interview. 

For the rushed scenario, a participant created gestures of 17--24 s of speech in a limited time of 5 min in a single trial. Each participant had eight trials with different speeches. A participant created gestures under both \textit{SGT} and \textit{MANUAL} conditions. One condition was used for the first four trials and the other condition for the next four trials. The order of the \textit{SGT} and \textit{MANUAL} conditions were balanced to avoid the order effects. Balancing was performed by groups of four trials rather than by individual trials. Half of the participants used SGT for the first four trials and MANUAL for the next four trials; the other participants used MANUAL first and SGT later. We chose this counterbalancing strategy from the concern that participants could be confused if they frequently switch the methods. The first two trials for each condition (1st, 2nd, 6th, and 7th trials) were for practices, and their results were not used in the later gesture quality evaluation. For the relaxed scenario, the lengths of speeches were 5--10 s, and a participant created gestures for two speeches for each condition (four trials in total). Short sentences were used in this scenario to ensure the participants have enough time to create gestures. The first trial of each condition (1st and 3rd trials) was for practice and its time was limited to 5 min. For the second trial, a participant had 20 min to create gestures and could finish it early if they were satisfied with the results or could not improve the gestures anymore. As a result, a participant in the rushed scenario created gestures for two speeches for each condition and a participant in the relaxed scenario made gestures for one speech for each condition, excluding practice trials.

We shuffled the order of speeches while satisfying two rules. First, there were no duplicated speeches throughout a participant's trials to avoid learning effects by authoring gestures for the same speech multiple times. Second, a single speech was included in both conditions to compare gesture quality between the conditions.

After the gesture design sessions, we asked participants whether they could author gestures as intended and were the created gestures appropriate to the speech for each condition. Participants answered on a Likert scale with a seven-point response format (strongly disagree -- strongly agree). In the interview, an experimenter asked the pros and cons of the \textit{SGT} and \textit{MANUAL} methods.

We recruited 12 participants from a university community for each scenario (24 participants in total) and prepared 12 speeches randomly sampled from the TED testset. The average ages of the participants for the rushed and relaxed scenario were 23.33 and 26.75, respectively. Four female and eight male people participated in both scenarios. The study was conducted remotely (due to COVID-19). The participants connected to a remote server that runs the SGToolkit. To control the experiment environment, we required participants to use a PC or laptop with a keyboard, a mouse, and a display larger than 13-inches. None of the participants had prior experience with gesture design. Although our toolkit is for both experts and non-experts, we thought the toolkit would have a bigger impact on non-experts by enabling easy and fast gesture authoring; thus, we conducted the user study with non-expert users.
 

\begin{figure}[t]
  \centering
  \includegraphics[width=0.8\linewidth]{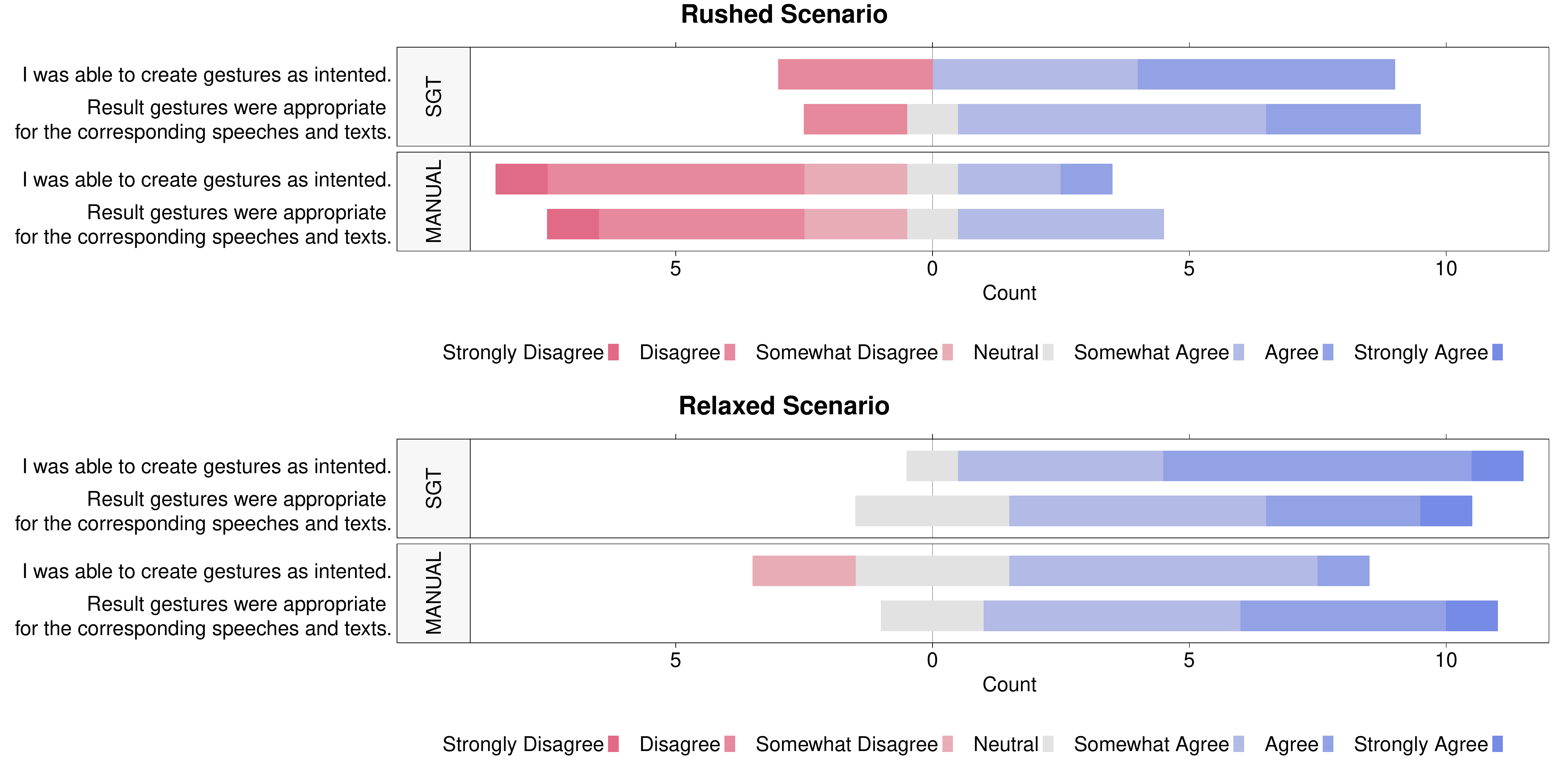}
  \caption{Survey results of the gesture design session for rushed and relaxed scenarios. We asked participants whether they could author gestures as intended and were the created gestures appropriate to the speech. A Likert scale with a seven-point response format (strongly disagree -- strongly agree) is used.}
  \label{fig:user_study_survey}
\end{figure}

\subsubsection{Results}

Figure~\ref{fig:user_study_survey} shows the survey results. In the rushed scenario, \textit{SGT} showed higher intention rating (M=4.7) and appropriateness rating (M=4.7) than \textit{MANUAL} (M=3.1 for intention, M=3.3 for appropriateness). This indicates that the proposed model helped the participants make gestures as intended. The participants also could create more suited gestures to the speeches with the proposed model than without it. In contrast, when they had enough time to create gestures (relaxed scenario), there were no significant differences between \textit{SGT} and \textit{MANUAL} in terms of intention and appropriateness. For statistical analyses, we used the Friedman test because none of the metrics passed the Shapiro-Wilk normality test. \textit{SGT} and \textit{MANUAL} showed a statistically significant difference for intention ($\chi^{2}_{F}(1)=9, p<0.05$) and appropriateness ($\chi^{2}_{F}(1)=8, p<0.05$) assessments only in the rushed scenario. 

We learned that the SGToolkit has its benefits and limitations from the interviews. The participants said it was easy to create gestures in the \textit{SGT} condition because they only had to add new gestures to some parts of the generated gestures. Moreover, automatically generated gestures acted as a guideline; the participants could get clues on what gestures to create. Additionally, the participants thought small and continuous movements of the generated gestures made the final output look more natural. They mentioned that it was hard to create those small movements without the generative model. 
However, while the \textit{SGT} condition showed a higher intention rating than \textit{MANUAL} on average, a few participants commented they had more control in the \textit{MANUAL} condition because the generative model did not fully adopt their pose and style controls. Also, a participant commented that starting from scratch in the \textit{MANUAL} condition was better because their imagination was not limited by a rough sketch.

We measured the completion time for the relaxed scenario. Only two participants used the whole 20 min in the \textit{MANUAL} condition; the others ended the gesture design task before the time limit. The average completion time (in s) was shorter in \textit{SGT} (M=760, SD=308) than \textit{MANUAL} (M=817, SD=276) but the difference was not statistically significant. The reason for no significant difference in time might be due to unclear terminating conditions. The participants kept trying to improve gesture quality as much as they could, so they might have used more time than needed for both conditions. We let the participants end the task when they are satisfied with the results. However, each participant had a different level of satisfaction, so this might have increased the deviation of the completion time. For the rushed scenario, all the participants used the whole 5 min.

\subsection{Subjective Gesture Quality Evaluation}

To assess the gesture quality of the SGToolkit, we conducted evaluation studies for the gestures generated from the gesture design study. The gestures were evaluated by other subjects recruited from Prolific, a crowdsourcing platform. We compared four conditions for the same speech: 1) reference human motion (\textit{REF}), 2) automatically generated gestures (\textit{AUTO}), 3) user-created gestures using the SGToolkit (\textit{SGT}), and 4) manually created gestures (\textit{MANUAL}). \textit{REF} was used for us to
know the referential quality of actual human gestures and \textit{AUTO} was for the quality of the automatic gesture generation without control inputs.

\subsubsection{Evaluation Method}
We followed the evaluation method used in the GENEA Challenge~\cite{geneaworkshop}, where different gesture generation models were compared. Video stimuli were rendered from gesture data by using the same visualization pipeline and a virtual character. Each evaluation page presented the videos of four conditions for the same speech, and a participant rated each video on a scale of 0--100. This evaluation method~\cite{jonell2021hemvip} was inspired by MUSHRA~\cite{itu2014method}, which is the standardized evaluation method for comparing audio qualities. We evaluated two different aspects of gestures as the GENEA Challenge did. First, we evaluated human-likeness, how much the gesture motion is human-like. The second one is the appropriateness, how well the gestures match speech acoustically and semantically. In human-likeness studies, the videos were muted to assess only the quality of the motion regardless of speech. And human-likeness and appropriateness were evaluated by different subjects to prevent participants from confusing evaluation criteria. There were two evaluating factors of human-likeness and appropriateness and two sets of results from the rushed and relaxed scenarios; thus, in total, we conducted four separate gesture quality assessment studies. 

Each participant rated 8 pages (32 videos in total, the rushed scenario) or 12 pages (48 videos in total, the relaxed scenario). We restricted the number of pages to 8 for the rushed scenario since the videos were longer than the ones in the relaxed scenario. For the condition \textit{SGT} and \textit{MANUAL} in the rushed scenario, two gesture design results are available per speech, so we selected one randomly for each participant. The presentation order of sentences and conditions were also randomized to reduce the order effect. 


\subsubsection{Participants}
Thirty subjects participated in each study; in total, there were 120 participants (18 to 67 years old, the average age is 35.9; 63 male, 56 female, 1 undisclosed). We restricted their current residency to English-speaking countries to ensure participants had no problem in understanding speech content (although 10 self-reported that they are non-native English speakers in the demographic survey). The median completion time was 21, 19, 16, and 16 min for the study of human-likeness in the rushed scenario, appropriateness in rushed, human-likeness in relaxed, and appropriateness in relaxed, respectively. Each participant was rewarded with 3.4 USD.

We used a similar attention check used in the GENEA challenge. We inserted attention checks in the two random pages per participant. The attention check video contained a text description asking to rate the current video with a certain number. The text shows up in the middle of the video, so the participants could not cheat by skimming video thumbnails. Who failed in the attention check was immediately rejected and not counted in the result analysis. Eighteen participants were rejected in total. 

\subsubsection{Results}

\begin{figure}[t]
  \centering
  \includegraphics[width=0.48\linewidth]{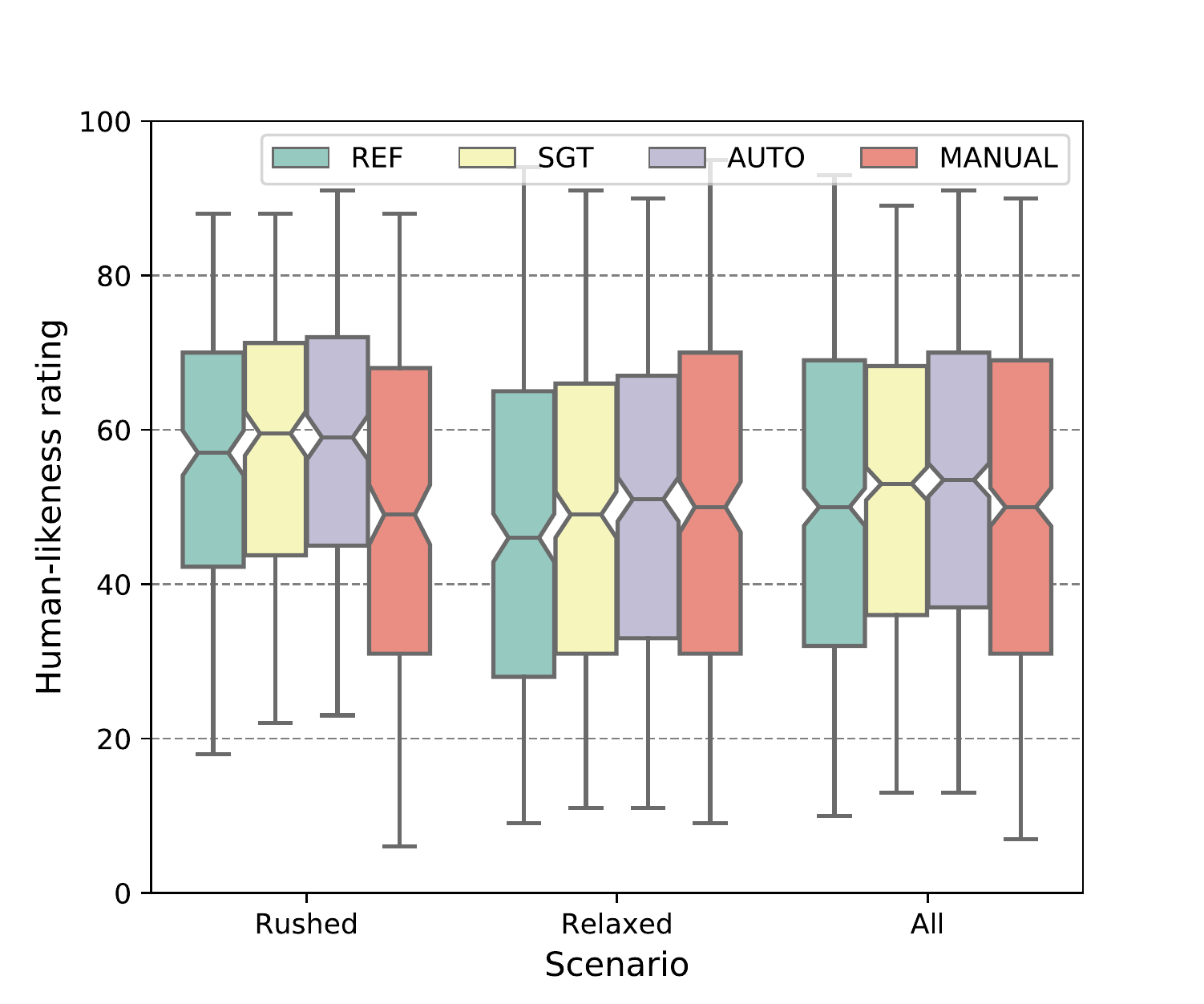} \includegraphics[width=0.48\linewidth]{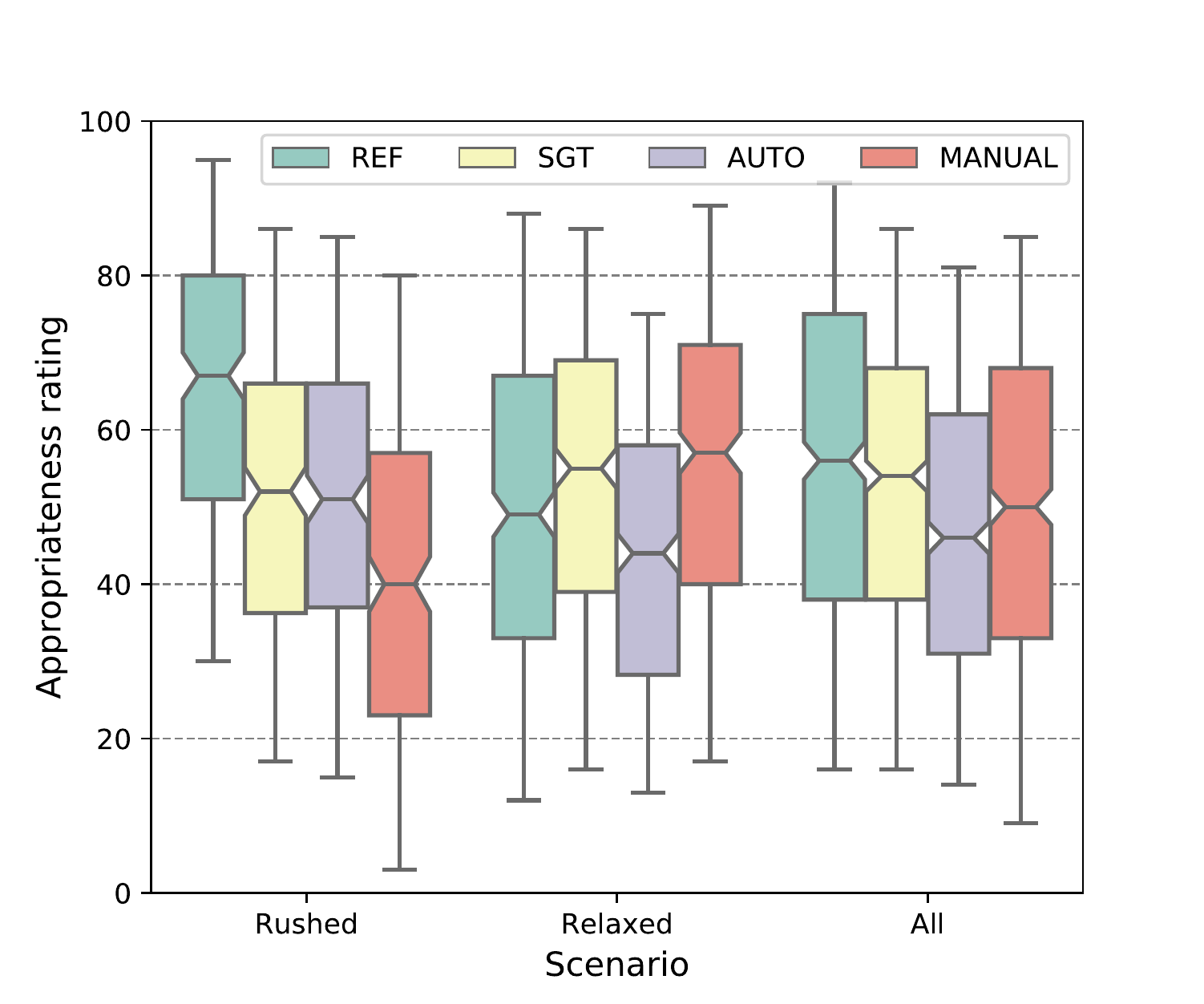}
  \caption{Gesture quality assessment results for (left) human-likeness and (right) appropriateness measures. Box plots for the four conditions are grouped by the scenario (rushed, relaxed, and both). Box notch represents a median value, and box boundaries represent the 25th and 75th percentiles. Whiskers represent the 5th and 95th percentiles.} \label{fig:qualityassessment}
\end{figure}

We collected 4560 ratings in total, and Figure \ref{fig:qualityassessment} visualizes the ratings using box plots for each condition and scenario. For the human-likeness study, all the conditions showed similar ratings, but only \textit{MANUAL} showed a significantly lower rating than the others in the rushed scenario ($Z$=-3.07, -3.87, -4.91 for comparing to \textit{REF, AUTO, SGT}, $p<0.001$ for all). We found more significant differences between conditions in the appropriateness study. For the rushed scenario, \textit{SGT} showed significantly higher ratings than \textit{MANUAL} ($Z$=6.83, $p<0.001$), but no significant difference to \textit{AUTO} condition ($Z$=0.72, $p$=0.47). For the relaxed scenario, \textit{SGT} and \textit{MANUAL} showed significantly higher ratings than \textit{AUTO} (Z=6.21, 6.52, $p<0.001$). When we consider appropriateness ratings from both scenarios, the best to worst rating order was \textit{REF} (median rating of 56), \textit{SGT} (54), \textit{MANUAL} (50), and \textit{AUTO} (46). \textit{REF} and \textit{SGT} showed significantly higher ratings than \textit{MANUAL} and \textit{AUTO} ($p<0.01$ for all). All the significances mentioned above were tested by Wilcoxon signed-rank tests with Holm-Bonferroni correction for multiple comparisons. We used the alpha value of 0.01 after the Holm-Bonferroni correction.

In order to see learning effects, we compared assessment scores of the gestures from the first and second trials (except practice trials) in the rushed scenario. We could not find significant differences between two trials in both \textit{SGT} ($Z$=0.19, $p$=0.84) and \textit{MANUAL} ($Z$=-0.30, $p$=0.76) from Wilcoxon rank-sum tests.

\section{Discussion} \label{sec:discussion}
We discuss the merits of the SGToolkit compared to the previous methods of keyframe animation and automatic gesture generation.

\textbf{SGToolkit vs. Manual keyframe animation.} The SGToolkit aimed to provide a more efficient way to author speech gestures than the manual keyframe animation by combining automatic generation and manual controls. We conducted the user study in two scenarios where the participants had enough time to create gestures and where they had a limited time. When the participants had enough time, the manual method showed the highest appropriateness rating (median rating=57) according to the gesture quality evaluation. However, when the production time is limited, the SGToolkit showed higher gesture quality than the manual method in terms of motion human-likeness (median ratings of 60 and 49) and gesture appropriateness (median ratings of 52 and 40). This result indicates that users can create gestures more efficiently using the SGToolkit. 
A few participants commented that they ran out of time because they had to insert keyframes tightly in the manual method to make a realistic motion. They also commented that less effort was required with the SGToolkit since it provides initial gestures and fills the non-controlled regions automatically.
The participants of the gesture design session for the rushed scenario reported that they could put their intention well with the SGToolkit (M=4.7, SD=1.7) than the manual method (M=3.1, SD=1.6) and also reported that they think the output gestures were more appropriate in the SGToolkit (M=4.7, SD=1.4) than the manual method (M=3.3, SD=1.5). 
    
\textbf{SGToolkit vs. Automatic generation.} The SGToolkit provides a unique function of pose and style controls over the automatic method. This control function was elicited from the expert interviews, and we integrated the control vectors into the neural generative model. The numerical validation showed that the proposed model successfully supports pose and style controls as well as maintaining automatic generation capability. The participants in the user study agreed that they could control gestures as they intended by using the SGToolkit (M=5.1 out of 7; average of two scenarios). 
Also, we found that the generated gestures using the SGToolkit (median rating=55) were more appropriate than the automatic method (median rating=44) in the relaxed scenario.
The SGToolkit did not outperform the automatic method in the rushed scenario, but it still has benefits for containing a designer's intention, which is an important requirement but not possible only with the automatic generation model.


We discussed the merits of the SGToolkit over two existing methods. Although the SGToolkit showed similar performance with \textit{AUTO} in the rushed scenario and \textit{MANUAL} in the relaxed scenario regarding the gesture quality, the SGToolkit has another merit of practicality. In a real-world situation, it is difficult for users to estimate the amount of gesture design work, and this makes the choice between existing \textit{AUTO} and \textit{MANUAL} methods difficult. With the SGToolkit, the user could start with automatic generation and a few controls when authoring time is limited, and they would add more controls on top of the previous authoring to refine gestures when a higher quality is required. We expect the SGToolkit to be a practical and go-to toolkit for gesture design.

We discuss whether the proposed SGToolkit meets the three requirements elicited from the expert interviews. 

\noindent
(\verb|Req.1|) \textbf{The toolkit should allow a user to alter the gesture motion by controlling one or more poses at specific time frames.}
We embedded a pose control in the neural generative model so that users can add a desired pose or gestures at a specific position. We validated qualitatively and quantitatively that the trained model can generate human-like gestures as following pose constraints specified by users. The participants in the user study also confirmed that they could put their intention as they wanted. When the production time was limited (the rushed scenario), the SGToolkit showed better controllability than the keyframe animation method.

\noindent
(\verb|Req.2|) \textbf{The toolkit should support modifying abstract aspects of gestures, such as style.} 
To enable users to control styles of gestures, we also integrated style controls in the model. With the style controls, users can adjust the overall speed, gesture space, and handedness of the output gestures. This style control only requires three scalar values, so it is handier to use than the pose controls which require a sequence of desired human poses. We also confirmed the proposed style control is working well through the numerical analysis.

\noindent
(\verb|Req.3|) \textbf{The toolkit should output gestures in a general format and provide platform-independent APIs so that users in different fields can use them.} The output gestures were represented as a sequence of 3D coordinates for each joint, so that any embodied agent can play the generated gestures with simple data conversion. We demonstrated playing gestures on Babylon.js (in the web client) and Blender (to render videos for the crowd-sourced gesture quality evaluation). For the Blender animation, we converted 3D joint coordinates to joint rotations by using Blender APIs. Users can export Blender animations to well-known animation formats such as BVH which can be used in other graphics and robotics systems. In addition, the SGToolkit is divided into a server to run the gesture generation model and a client to interact with users, and the server and client communicate by using REST APIs. Users can develop other clients dedicated to a specific platform easily.

Finally, we discuss the limitations of the SGToolkit. First, a few participants reported that output gestures are not perfectly the same as pose controls. The difference was small but noticeable. This happens especially when unrealistic poses were inputted as pose controls. We would add optional post-processing which blends the output gestures and the pose controls for users who want precise controls. 
Second, the toolkit lacks facial and hand motion. Extending the current model to synthesize facial, arm, and hand motion altogether might be a solution, but it is not feasible without a proper dataset. Another solution is to generate the motions one by one and integrate them into an embodied agent. Facial motion is independent to some extent of gestures, so it is possible to integrate facial motions and gestures. For hands, because hands (hand orientation and finger motion) are correlated to arm gestures, we would not synthesize them independently. Instead, a recent study tried inferring hand motion from arm gestures and they showed promising results \cite{ng2021body2hands}. A possible future direction is to integrate a hand motion generator as well as a facial motion generator into the SGToolkit for completeness and convenience for users.

\section{Conclusion} \label{sec:conclusion}

In this paper, we presented the SGToolkit, an open-sourced and platform-agnostic toolkit to author speech gestures. From the requirements elicited from expert interviews, we designed a generative neural model that can generate speech gestures automatically and also accommodates users' intent described as pose and style controls. Thus, the proposed toolkit has both merits of automatic gesture generation and manual authoring. Users could author gestures easily by having a rough sketch automatically generated and add controls as they want on top of the sketch. Usability and gesture qualities were evaluated through a user study and a crowd-sourced assessment. The results showed that the SGToolkit gives unique advantages over both the manual authoring and previous automatic generation method. The public availability of the toolkit would be beneficial to designers, developers, and researchers related to embodied conversational agents. 

\begin{acks}
This work was supported by the Institute of Information \& communications Technology Planning \& Evaluation (IITP) grant funded by the Korea government (MSIT) (No. 2017-0-00162, Development of Human-care Robot Technology for Aging Society, 70\%; No. 2020-0-00537, Development of 5G based low latency device –- edge cloud interaction technology, 30\%). 
Resources supporting this work were provided by the Ministry of Science and ICT and NIPA (`HPC Support' Project).
\end{acks}

\bibliographystyle{ACM-Reference-Format}
\bibliography{main}

\appendix
\section{Implementation Details}
\subsection{Generating Gestures for a Long Speech} \label{appendix:long}
The model outputs a sequence of 30 poses for a single run. This length is fixed at both training and test time. For a long sequence generation, we ran the model several times for different chunks of speech. The sliding chunks span 60 frames and they overlap 30 frames. This sliding window strategy is the same as the previous study~\cite{yoon2020speech}. Also, we used the last 30 frames of the last chunk as pose controls in the gesture generation for the next chunk to make smooth transitions between consecutive chunks. Please refer to the codes for details.

\subsection{Training Details}
Each pose was represented as a collection of 10 upper-body joints of nose, head-top, neck, spine, R/L shoulders, R/L elbows, and R/L wrists. We used this representation in the paper for better understanding; however, in the actual implementation, we used directional vectors of adjacent joints for faster and stable training. Speech words were represented as one-hot vectors indicating a word position in the dictionary which was built on the training set of the TED dataset.

\begin{figure}[t]
  \centering
  \includegraphics[width=0.6\linewidth]{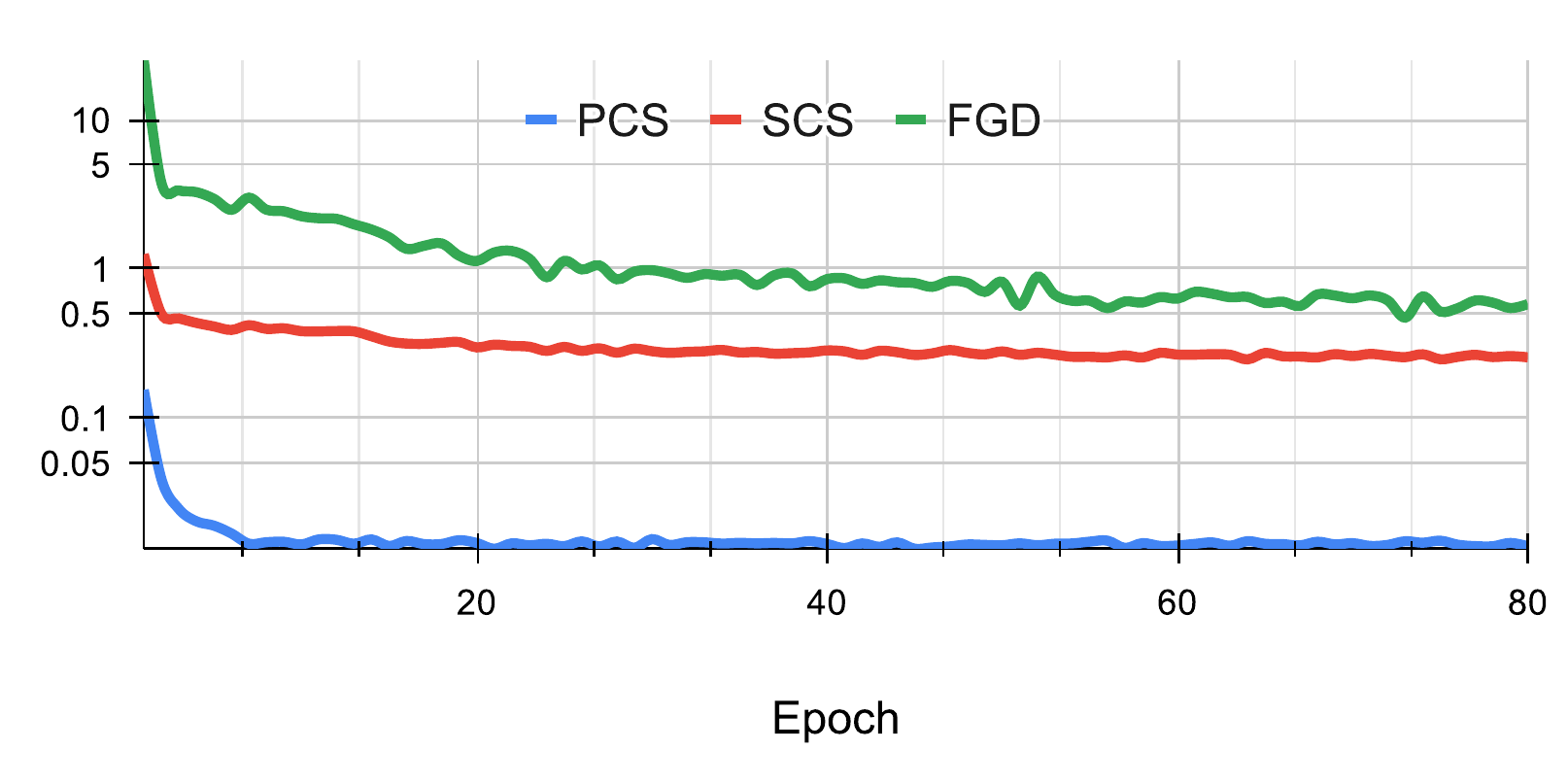} 
  \caption{Validation curves for the gesture generation model used in the SGToolkit. The curves show PCS, SCS, and FGD trends during 80 epochs of training. The y-axis is in log scale.} \label{fig:validation_curve}
\end{figure}

The model was trained for 80 epochs using the ADAM optimizer with the learning rate of 0.0005. The mini-batch size was 128. Figure \ref{fig:validation_curve} shows the validation curves showing FGD, PCS, SCS on the validation dataset during 80 training epochs.

\end{document}

%% file: header1.tex
\copyrightyear{2021}
\acmYear{2021}
\setcopyright{licensedothergov}\acmConference[UIST '21]{The 34th Annual ACM Symposium on User Interface Software and Technology}{October 10--14, 2021}{Virtual Event, USA}
\acmBooktitle{The 34th Annual ACM Symposium on User Interface Software and Technology (UIST '21), October 10--14, 2021, Virtual Event, USA}
\acmPrice{15.00}
\acmDOI{10.1145/3472749.3474789}
\acmISBN{978-1-4503-8635-7/21/10}

\acmSubmissionID{3346}

%% file: header2.tex

\title{SGToolkit: An Interactive Gesture Authoring Toolkit for Embodied Conversational Agents}

\author{Youngwoo Yoon}
\authornote{Both authors contributed equally to this research.}
\affiliation{%
  \institution{ETRI \& KAIST}
  \country{Republic of Korea}
}
\email{youngwoo@etri.re.kr}
\author{Keunwoo Park}
\authornotemark[1]
\affiliation{%
  \institution{KAIST}
  \country{Republic of Korea}
}
\author{Minsu Jang}
\affiliation{%
  \institution{ETRI}
  \country{Republic of Korea}
}
\author{Jaehong Kim}
\affiliation{%
  \institution{ETRI}
  \country{Republic of Korea}
}
\author{Geehyuk Lee}
\affiliation{%
  \institution{KAIST}
  \country{Republic of Korea}
}


\begin{abstract}
Non-verbal behavior is essential for embodied agents like social robots, virtual avatars, and digital humans. Existing behavior authoring approaches including keyframe animation and motion capture are too expensive to use when there are numerous utterances requiring gestures. Automatic generation methods show promising results, but their output quality is not satisfactory yet, and it is hard to modify outputs as a gesture designer wants. We introduce a new gesture generation toolkit, named SGToolkit, which gives a higher quality output than automatic methods and is more efficient than manual authoring. For the toolkit, we propose a neural generative model that synthesizes gestures from speech and accommodates fine-level pose controls and coarse-level style controls from users. The user study with 24 participants showed that the toolkit is favorable over manual authoring, and the generated gestures were also human-like and appropriate to input speech. The SGToolkit is platform agnostic, and the code is available at \url{https://github.com/ai4r/SGToolkit}.
\end{abstract}

\ccsdesc[500]{Human-centered computing~Interaction techniques}
\ccsdesc[300]{Human-centered computing~Interactive systems and tools}
\ccsdesc[300]{Computer systems organization~Robotics}

\keywords{speech gestures, social behavior, gesture authoring, toolkit}

\maketitle